\documentclass{sf2a-conf}
\usepackage{graphicx}
\newcommand{\CD}{{\cal D}}
\newcommand{\CR}{{\cal R}}
\newcommand{\average}[1]{\left\langle #1 \right\rangle_\CD}

\newcommand{\initial}[1]{{#1_{\rm \bf i}}}

\begin{document}

\TitreGlobal{SF2A 2006}

\title{Reinterpreting quintessential dark energy through averaged inhomogeneous cosmologies}
\author{Julien Larena}\address{Laboratoire de l'Univers et de ses Th\'eories
  (LUTH), CNRS UMR 8102, Observatoire de Paris and Universit\'e Paris 7 Denis
  Diderot}
\author{Thomas Buchert}\address{Fakult\"at f\"ur Physik, Universit\"at Bielefeld, Postfach 100131,
33501 Bielefeld and, Arnold Sommerfeld Center for Theoretical Physics ASC,
Ludwig--Maximilians--Universit\"{a}t, Theresienstra{\ss}e 37,
80333 M\"{u}nchen, Germany}
\author{Jean-Michel Alimi$^{1}$}
\runningtitle{Correspondence between quintessence and backreaction}
\setcounter{page}{237}
\index{Julien Larena}
\index{Thomas Buchert}
\index{Jean-Michel Alimi}
\maketitle
\begin{abstract}
Regionally averaged relativistic cosmologies have recently been considered as
a possible explanation for the apparent late time acceleration of the
Universe. This contribution reports on a mean field description of the
backreaction in terms of a minimally coupled regionally homogeneous scalar
field evolving in a potential, then giving a physical origin
to the various phenomenological scalar fields generically called quintessence
fields. As an example, the correspondence is then applied to scaling solutions. 
\end{abstract}
%
\section{Introduction}

Modern cosmology is nowadays settled on observations concerning mainly the
distribution of matter and the dynamics of the expansion of
the Universe.
On the one hand, there are now various cosmological observations supporting a matter
distribution that is homogeneous on large scales of order $100$ h$^-1$Mpc. Nevertheless, at late times, the matter distribution is highly
structured on smaller scales, with the presence of clusters of galaxies,
filaments and voids. Moreover, the statistical isotropy of the Cosmic
Microwave Background radiation supports the idea that the Universe is
highly isotropic on average, on large scales. 
Facing these observational issues, one assumes in the standard cosmological
framework that the Universe is homogeneous and isotropic on all scales,
resulting in a spacetime described by a Friedmann-Robertson-Walker (FRW)
metric, the inhomogeneities being perturbations around this homogeneous and
isotropic background. Then, all the observables of the Universe on large scales
can be deduced from a single degree of freedom: the scale factor of the metric, and
the dynamics of the inhomogeneities is well described as long as the density contrast in the matter fields remains small.

On the other hand, many recent observations strongly favor a Universe whose
expansion has been accelerating in the recent past and may be accelerating
today. In the FRW context this necessarily requires the introduction of
exotic sources as for example a cosmological constant or quintessence fields,
or a modification of gravity, generally implying the so-called coincidence
problem: why is the expansion accelerating approximatelly at the same time
when the Universe becomes structured, that is when the density contrast in the
matter field is no longer small on a wide range of scales? 

Regionally averaged relativistic cosmologies may be able to answer this
question by linking the dynamics of the Universe on large scales to its
structuration on smaller scales; see interesting discussions of that topic in
R\"as\"anen (2006). It consists in defining cosmologies that are
homogeneous on large scales without supposing any local symmetry, thanks to a
spatial averaging procedure. It results in equations for a volume scale factor
that not only include an averaged matter source term, but also additional terms that can
be interpreted as the effects of the coarse-grained inhomogeneities on the
large scales dynamics. These additional terms are commonly named backreaction.

In this paper, after introducing the formalism of regionally averaged
cosmologies in the first part, we shall propose a correspondence between
regionally averaged cosmologies and Friedmannian scalar field cosmologies in
the second part, the scalar field being interpreted in this context as a mean
field description of the inhomogeneous Universe, that can play the role of a
quintessence field.
Then, in the third part, as an example of the correspondence, we explicitly reconstruct the mean field theory for the
particular class of scaling solutions of the regionally averaged cosmologies,
and discuss its properties. This correspondence has been proposed and
discussed in Buchert et al. (2006).

\section{Regionally averaged cosmologies: the backreaction context}

In this paper, since we are interested in the late time behavior of the
cosmological model, we restrict the analysis to a Universe filled with an
irrotational fluid of dust matter with density $\rho (t,X^{i})$. The more general case of an irrotational
perfect fluid can be found in Buchert (2001).

\subsection{Averaged ADM equations}

Following Buchert (2000) we foliate the spacetime by flow-orthogonal hypersurfaces with the
3-metric $g_{ij}$. The line element then reads $ds^{2}=-dt^{2}+g_{ij}dX^{i}dX^{j}$.
The large scale homogeneous model is built by averaging the scalar part of the general
relativistic equations on a spatial domain $\CD$ with a spatial averager
applied to any scalar function $\Upsilon (t,X ^{i})$:
\begin{equation}
\label{averager}
\average{\Upsilon (t,X ^{i})}=\frac{1}{V_{\CD}}\int_{\CD}\Upsilon (t,X
^{i})Jd^{3}X \mbox{ ,}
\end{equation}
where $V_{\CD}$ is the volume of the domain $\CD$ and $J=\sqrt{\mbox{det}
  (g_{ij})}$.
Then, one can define a volume scale factor
  $a_{\CD}=(V_{\CD}/V_{\initial\CD})^{1/3}$, and applying the averager
  \ref{averager} to the Hamiltonian constraint and Raychaudhuri's equation when $\Lambda$ has been set to $0$ leads to :
\begin{eqnarray}
\label{averagedsystem}
\left(\frac{\dot{a}_{\CD}}{a_{\CD}}\right)^{2}&=&\frac{8\pi
  G}{3}\average{\rho}-\frac{\average{\CR}+{\cal Q}_{\CD}}{6}\nonumber\\
\frac{\ddot{a}_{\CD}}{a_{\CD}}&=&-\frac{4\pi G}{3}\average{\rho}+\frac{{\cal
    Q}_{\CD}}{3}\\
a_{\CD}^{-6}\partial_{t}\left(a_{\CD}^{6}{\cal
  Q}_{\CD}\right)&=&-a_{\CD}^{-2}\partial_{t}\left(a_{\CD}^{2}\average{\CR}\right)\nonumber
  \mbox{ ,}
\end{eqnarray}
where $\average{\CR}$ is the averaged spatial 3-Ricci scalar, and ${\cal
  Q}_{\CD}$ is known as the kinematical backreaction term. This backreaction
  is given in terms of the well-known ADM variables that are the local expansion rate $\theta$ and the rate of shear
  $\sigma$ by: ${\cal Q}_{\CD}=2\average{(\theta-\average{\theta})^{2}}/3-2\average{\sigma^{2}}$.
  One can notice that this additionnal term is the spatial variance over the domain
  $\CD$ of these quantities. In other words, the more the matter distribution
  is structured, with collapsed regions and voids, the more this term may contribute to the dynamics, except of
  course if the two parts, i.e. expansion and shear fluctuations compensate
  each other.
The third equation of the system \ref{averagedsystem} is simply an
  integrability condition that expresses the compatibility of the first two equations.

\subsection{Large scale homogeneous model}

The system \ref{averagedsystem} characterizes the properties of the Universe
on large scales. It preserves the main feature of the standard FRW Universe,
that is the fact that the properties of the Universe on large scales can be
deduced from a single scale factor, but this scale factor now obeys dynamical
equations that differ from the FRW equations for a dust field because of the
additional source terms ${\cal Q}_{\CD}$ and $\average{\CR}$. These terms
arise because the averaging and the time derivatives don't commute. Of course
the curvature is also present in FRW equations, but it reduces to a constant
curvature term, whereas in averaged cosmologies, it is coupled to the
backreaction term through the last equation of \ref{averagedsystem}. We will
see below that this coupling is essential to explain the cosmic acceleration
in averaged cosmologies.
In analogy with FRW cosmology, we introduce $H_{\CD}=\dot{a}_{\CD}/a_{\CD}$,
and we can define a set of cosmological parameters:
\begin{equation}
\Omega_{m}^{\CD}=8\pi G\average{\rho}/3H_{\CD}^{2}\mbox{ , }\Omega_{\CR}^{\CD}=-\average{\CR}/6H_{\CD}^{2}
\mbox{ , }\Omega_{\cal Q}^{\CD}=-{\cal Q}_{\CD}/6H_{\CD}^{2}
\end{equation}
as well as an effective deceleration parameter:
$q_{\CD}=-\ddot{a}_{\CD}/(a_{\CD}H_{\CD}^{2})=\Omega_{m}^{\CD}/2+2\Omega_{\cal
  Q}^{\CD}$.

To emphasize the difference between the mean curvature $\average{\CR}$ and the
Friedmannian constant curvature $-k/a_{\CD}^{2}$, one should note that they
differ by a term representing the effect of the whole history of the Universe
since the beginning of the dust dominated phase: $k/a_{\CD}^{2}=(\average{\CR}+{\cal Q}_{\CD})/6+\frac{2}{3a_{\CD}^{2}}\left(\int_{1}^{a_{\CD}}a{\cal Q}_{\CD}(a)da\right)$.

So, when the backreaction term doesn't vanish identically, the mean curvature
doesn't behave like a constant curvature term.
Finally, it is important to note that the system \ref{averagedsystem} is not
closed: it has four unknown quantities, but only three independent
equations. In order to close it, it is then necessary to introduce another
relation that can be either a mathematical ansatz or a physical statement.

\section{Correspondence with scalar field cosmologies}

In order to constrain and understand the dynamics of averaged cosmologies, it
could be interesting to benefit from the well-known properties of the
Friedmannian cosmologies, so we will develop in this section a correspondence between the
backreaction effect, and the simplest mean field model, that is a homogeneous
minimally coupled scalar field $\Phi_{\CD}(t)$ with a self-interaction
potential $U(\Phi_{\CD})$. Let's parameterize ${\cal Q}_{\CD}$ and $\average{\CR}$
as follows:
\begin{equation}
-\frac{1}{8\pi G}{\cal Q}_{\CD}=\epsilon
 \dot{\Phi}_{\CD}^{2}-U(\Phi_{\CD})\mbox{ , }-\frac{1}{8\pi G}\average{\CR}=3
 U(\Phi_{\CD})\mbox{ ,}
\end{equation}
where $\epsilon=+1$ for a standard scalar field and $\epsilon=-1$ for a
phantom scalar field. Then, the system \ref{averagedsystem} becomes:
\begin{eqnarray}
\label{scalarsystem}
\left(\frac{\dot{a}_{\CD}}{a_{\CD}}\right)^{2}&=&\frac{8\pi
  G}{3}\left(\average{\rho}+\frac{\epsilon}{2}\dot{\Phi}_{\CD}^{2}+U(\Phi_{\CD})\right)\nonumber\\
\frac{\ddot{a}_{\CD}}{a_{\CD}}&=&-\frac{4\pi
  G}{3}\left(\average{\rho}+2\epsilon\dot{\Phi}_{\CD}^{2}-2U(\Phi_{\CD})\right)\\
\ddot{\Phi}_{\CD}+3\frac{\dot{a}_{\CD}}{a_{
  \CD}}\dot{\Phi}_{\CD}+\epsilon\frac{\partial U(\Phi_{\CD})}{\partial \Phi_{\CD}}&=&0\nonumber
\end{eqnarray}
Except for the dependence on the domain $\CD$, these are exactly the equations
for a homogeneous cosmology in presence of a dust field and a minimally
coupled scalar field. Because this scalar field appears as a mean field
description of the morphology of the structures in the Universe, we call it
the 'morphon field'.

\section{Example: the scaling solutions}

\subsection{The solutions}
The correspondence established in the previous section can be used in two
different ways. The first one, and probably the more useful would be to
consider particular models of scalar field cosmologies and to deduce the
characteristics of the corresponding backreaction and mean curvature; this
will be the subject of a forthcoming work. In this short contribution, as an
illustration of the correspondence, we will conversely focus on constructing
the mean field model from a particular class of backreaction.
We consider the large class of scaling solutions:
\begin{equation}
\label{scalingback}
{\cal Q}_{\CD}={\cal Q}_{\initial\CD} a_{\CD}^{n} \mbox{ , } \average{\CR}=\CR_{\initial \CD}a_{\CD}^{p}
\end{equation}    
where ${\cal Q}_{\initial\CD}$ and $\CR_{\initial \CD}$ are the initial values
of ${\cal Q}_{\CD}$ and $\average{\CR}$, and $n$ and $p$ are real numbers.
Inserting this ansatz in the third equation of \ref{averagedsystem}, one
obtains two different kinds of solutions. For $n\neq p$, the only solution is:
\begin{equation}
\label{ndiffp}
{\cal Q}_{\CD}={\cal Q}_{\initial\CD} a_{\CD}^{-6}\mbox{ , } \average{\CR}=\CR_{\initial \CD}a_{\CD}^{-2}
\end{equation}
that is a near-Friedmannian solution because it reduces to a constant
curvature for $a_{\CD} \rightarrow + \infty$. It corresponds to the case
where the backreaction and the mean curvature evolve independently.
On the contrary, the solutions for $n=p$: 
\begin{equation}
\label{neqp}
{\cal Q}_{\CD}=r\average{\CR}=r\CR_{\initial \CD}a_{\CD}^{n}\mbox{ , with } n=-2\frac{1+3r}{1+r}
\end{equation}
entail a strong coupling between the backreaction and the mean curvature. This
case is an extreme one, but the coupling must be considered a generic
property. The parameter $r$ that is constant for the scaling solutions is the
conversion rate of the mean curvature into backreaction; it plays a very
important role in the mechanism responsible for the cosmic acceleration: it
only occurs if $r\in ]0,-1[$, that is if the mean curvature converts sufficiently
    into backreaction that then decays slowly enough or even
    grows ($n>-2$). In the following we will focus on this class of strongly coupled solutions \ref{neqp}.

\subsection{Reconstruction of the associated morphon field}

One can then reconstruct the potential for the scalar field associated with
the scaling solutions \ref{neqp}. A straightforward calculation provides:
\begin{equation}
\label{pot}
U(\Phi_{\CD})=\frac{-(1+r)\CR_{\initial\CD}}{24\pi G}\left(\frac{-(1+r)\CR_{\initial\CD}}{16\pi G
  \langle\varrho\rangle_{\initial\CD}}\right)^{2\frac{1+3r}{1-3r}}\sinh^{-4\frac{1+3r}{1-3r}}\left(\frac{(1-3r)\sqrt{\pi
  G}}{\sqrt{\epsilon (1+3r)(1+r)}}\Phi_{\CD}\right) \mbox{ .}
\end{equation}

The potential \ref{pot} is known in
    the literature about scalar field type dark energy (Sahni et al. 2000,
    Sahni \& Starobinskii 2003, Urena-Lopez \& Matos 2000). It corresponds
    to dark energy with a constant equation of state, here given in terms of
    the parameters of the averaged cosmology by $w_{DE}=-(n+3)/3$.

\begin{figure}[h]
   \centering
   \includegraphics[width=6cm]{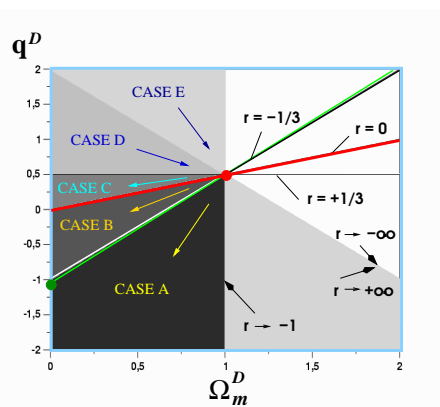}
      \caption{Phase space of the scaling solutions. Each scaling is a straight
      line passing through the Einstein-de Sitter model $(-1,1/2)$.}
       \label{fig:cosmicstates}
   \end{figure}

This result is consistent only under some restrictions on the parameters: we
must have $\CR_{\initial \CD}>0$ and $\epsilon=+1$ for $r<-1$ and
$\CR_{\initial \CD}<0$ for $r>-1$ with $\epsilon=-1$ in $r\in ]-1,-1/3[$ and
    $\epsilon=+1$ if $r>-1/3$. All these conditions show that scaling
    backreaction can reproduce a wide variety of cosmological scalar fields
    such as standard quintessence, phantom quintessence, a cosmological
    constant. They can be classified as 'cosmic states' in the phase diagram $(\Omega_{m}^{\CD},q_{\CD})$
of figure \ref{fig:cosmicstates}. Case A are phantom dark energy models; cases B and
C are standard scalar field models in a decreasing potential; case D are standard
scalar fields in a well-type potential, and case E are standard scalar fields
rolling in a negative potential that is not bounded from below. The green line
represents the scalar field model inferred from the SNLS best fit model with
$w_{DE}=-1.02$ (Astier et al. 2006). The arrows in each sector represent  the evolution of the
models in time: the Einstein-de Sitter model $(1,1/2)$
appears as a saddle point for the dynamics.

\section{Conclusion}

The mean field description of backreaction effects through a scalar field does
not only provide a rephrasing of the kinematics of backreaction, but it also
justifies the existence of the cosmological effective scalar field, that may be
responsible for the cosmic acceleration, on the basis of an underlying
fundamental theory, that is Einstein General Relativity: the cosmic
quintessence emerges in the process of interpreting the real Universe in a
homogeneous context. 
The study of scaling solutions allowed to understand that
the cosmic acceleration is only possible if the mean curvature is strongly
coupled to the backreaction and converts into it to maintain it at a high
level. Nevertheless, more realistic solutions, with a varying conversion rate
must be investigated.
Finally, in order to firmly establish that backreaction effects are the source
of the acceleration of the expansion, it will be necessary to explicitly
compute these effects from generic relativistic models and observations of the large-scale structures of the Universe.




\begin{thebibliography}{}
\bibitem{}Astier P. \& al 2006, A\&A, 447, 31
\bibitem{}Buchert T. 2000, Gen. Rel. Grav., 32, 105
\bibitem{}Buchert T. 2000, Gen. Rel. Grav., 33, 1381
\bibitem{}Buchert T., \& Larena J., \& Alimi J. M. 2006, gr-qc/0606020
\bibitem{}R\"as\"anen S. 2006, astro-ph/0607626
\bibitem{}Sahni V., \& Starobinskii A. A. 2000, Int. J. Mod. Phys. D, 9, 373
\bibitem{}Sahni V., \& Saini T. D., \& Starobinskii A. A., \& Alam U. 2003,
  JETP Lett., 77, 201
\bibitem{}Urena-Lopez, \& Matos T. 2000, Phys. Rev. D, 62, 081302
\end{thebibliography}
\end{document}